\catcode`@=11 
%
%
%

\font\fourteenrm=cmr10 scaled\magstep2
\font\twelverm=cmr12

\font\ninerm=cmr9

\font\sixrm=cmr6

\font\fourteenbf=cmbx10 scaled\magstep2
\font\twelvebf=cmbx10 scaled\magstep1
\font\ninebf=cmbx9	      \font\sixbf=cmbx6
\font\seventeeni=cmmi10 scaled\magstep3	    \skewchar\seventeeni='177
\font\fourteeni=cmmi10 scaled\magstep2	    \skewchar\fourteeni='177
\font\twelvei=cmmi10 scaled\magstep1	    \skewchar\twelvei='177
\font\ninei=cmmi9			    \skewchar\ninei='177
\font\sixi=cmmi6			    \skewchar\sixi='177
\font\seventeensy=cmsy10 scaled\magstep3    \skewchar\seventeensy='60
\font\fourteensy=cmsy10 scaled\magstep2	    \skewchar\fourteensy='60
\font\twelvesy=cmsy10 scaled\magstep1	    \skewchar\twelvesy='60
\font\ninesy=cmsy9			    \skewchar\ninesy='60
\font\sixsy=cmsy6			    \skewchar\sixsy='60

\font\fourteenex=cmex10 scaled\magstep2
\font\twelveex=cmex10 scaled\magstep1

\font\fourteensl=cmsl10 scaled\magstep2
\font\twelvesl=cmsl10 scaled\magstep1

\font\ninesl=cmsl9

\font\fourteenit=cmti10 scaled\magstep2
\font\twelveit=cmti10 scaled\magstep1
\font\twelvett=cmtt10 scaled\magstep1

%

%
\font\twelvecp=cmcsc10 scaled\magstep1
\font\tencp=cmcsc10
\newfam\cpfam
%
%
\newcount\f@ntkey	     \f@ntkey=0
\def\samef@nt{\relax \ifcase\f@ntkey \rm \or\oldstyle \or\or
	 \or\it \or\sl \or\bf \or\tt \or\caps \fi }
\def\fourteenpoint{\relax
    \textfont0=\fourteenrm	    \scriptfont0=\tenrm
    \scriptscriptfont0=\sevenrm
     \def\rm{\fam0 \fourteenrm \f@ntkey=0 }\relax
    \textfont1=\fourteeni	    \scriptfont1=\teni
    \scriptscriptfont1=\seveni
     \def\oldstyle{\fam1 \fourteeni\f@ntkey=1 }\relax
    \textfont2=\fourteensy	    \scriptfont2=\tensy
    \scriptscriptfont2=\sevensy
    \textfont3=\fourteenex     \scriptfont3=\fourteenex
    \scriptscriptfont3=\fourteenex
    \def\it{\fam\itfam \fourteenit\f@ntkey=4 }\textfont\itfam=\fourteenit
    \def\sl{\fam\slfam \fourteensl\f@ntkey=5 }\textfont\slfam=\fourteensl
    \scriptfont\slfam=\tensl
    \def\bf{\fam\bffam \fourteenbf\f@ntkey=6 }\textfont\bffam=\fourteenbf
    \scriptfont\bffam=\tenbf	 \scriptscriptfont\bffam=\sevenbf
    \def\tt{\fam\ttfam \twelvett \f@ntkey=7 }\textfont\ttfam=\twelvett
    \h@big=11.9\p@{} \h@Big=16.1\p@{} \h@bigg=20.3\p@{} \h@Bigg=24.5\p@{}
    \def\caps{\fam\cpfam \twelvecp \f@ntkey=8 }\textfont\cpfam=\twelvecp
    \setbox\strutbox=\hbox{\vrule height 12pt depth 5pt width\z@}
    \samef@nt}
\def\twelvepoint{\relax
    \textfont0=\twelverm	  \scriptfont0=\ninerm
    \scriptscriptfont0=\sixrm
     \def\rm{\fam0 \twelverm \f@ntkey=0 }\relax
    \textfont1=\twelvei		  \scriptfont1=\ninei
    \scriptscriptfont1=\sixi
     \def\oldstyle{\fam1 \twelvei\f@ntkey=1 }\relax
    \textfont2=\twelvesy	  \scriptfont2=\ninesy
    \scriptscriptfont2=\sixsy
    \textfont3=\twelveex	  \scriptfont3=\twelveex
    \scriptscriptfont3=\twelveex
    \def\it{\fam\itfam \twelveit \f@ntkey=4 }\textfont\itfam=\twelveit
    \def\sl{\fam\slfam \twelvesl \f@ntkey=5 }\textfont\slfam=\twelvesl
    \scriptfont\slfam=\ninesl
    \def\bf{\fam\bffam \twelvebf \f@ntkey=6 }\textfont\bffam=\twelvebf
    \scriptfont\bffam=\ninebf	  \scriptscriptfont\bffam=\sixbf
    \def\tt{\fam\ttfam \twelvett \f@ntkey=7 }\textfont\ttfam=\twelvett
    \h@big=10.2\p@{}
    \h@Big=13.8\p@{}
    \h@bigg=17.4\p@{}
    \h@Bigg=21.0\p@{}
    \def\caps{\fam\cpfam \twelvecp \f@ntkey=8 }\textfont\cpfam=\twelvecp
    \setbox\strutbox=\hbox{\vrule height 10pt depth 4pt width\z@}
    \samef@nt}
\def\tenpoint{\relax
    \textfont0=\tenrm	       \scriptfont0=\sevenrm
    \scriptscriptfont0=\fiverm
    \def\rm{\fam0 \tenrm \f@ntkey=0 }\relax
    \textfont1=\teni	       \scriptfont1=\seveni
    \scriptscriptfont1=\fivei
    \def\oldstyle{\fam1 \teni \f@ntkey=1 }\relax
    \textfont2=\tensy	       \scriptfont2=\sevensy
    \scriptscriptfont2=\fivesy
    \textfont3=\tenex	       \scriptfont3=\tenex
    \scriptscriptfont3=\tenex
    \def\it{\fam\itfam \tenit \f@ntkey=4 }\textfont\itfam=\tenit
    \def\sl{\fam\slfam \tensl \f@ntkey=5 }\textfont\slfam=\tensl
    \def\bf{\fam\bffam \tenbf \f@ntkey=6 }\textfont\bffam=\tenbf
    \scriptfont\bffam=\sevenbf	   \scriptscriptfont\bffam=\fivebf
    \def\tt{\fam\ttfam \tentt \f@ntkey=7 }\textfont\ttfam=\tentt
    \def\caps{\fam\cpfam \tencp \f@ntkey=8 }\textfont\cpfam=\tencp
    \setbox\strutbox=\hbox{\vrule height 8.5pt depth 3.5pt width\z@}
    \samef@nt}
%
%
%
%
\newdimen\h@big  \h@big=8.5\p@
\newdimen\h@Big  \h@Big=11.5\p@
\newdimen\h@bigg  \h@bigg=14.5\p@
\newdimen\h@Bigg  \h@Bigg=17.5\p@
\def\big#1{{\hbox{$\left#1\vbox to\h@big{}\right.\n@space$}}}
\def\Big#1{{\hbox{$\left#1\vbox to\h@Big{}\right.\n@space$}}}
\def\bigg#1{{\hbox{$\left#1\vbox to\h@bigg{}\right.\n@space$}}}
\def\Bigg#1{{\hbox{$\left#1\vbox to\h@Bigg{}\right.\n@space$}}}
%
%
%
\normalbaselineskip = 20pt plus 0.2pt minus 0.1pt
\normallineskip = 1.5pt plus 0.1pt minus 0.1pt
\normallineskiplimit = 1.5pt
\newskip\normaldisplayskip
\normaldisplayskip = 20pt plus 5pt minus 10pt
\newskip\normaldispshortskip
\normaldispshortskip = 6pt plus 5pt
\newskip\normalparskip
\normalparskip = 6pt plus 2pt minus 1pt
\newskip\skipregister
\skipregister = 5pt plus 2pt minus 1.5pt
\newif\ifsingl@	   \newif\ifdoubl@
\newif\iftwelv@	   \twelv@true
\def\singlespace{\singl@true\doubl@false\spaces@t}
\def\doublespace{\singl@false\doubl@true\spaces@t}
\def\normalspace{\singl@false\doubl@false\spaces@t}
\def\Tenpoint{\tenpoint\twelv@false\spaces@t}
\def\Twelvepoint{\twelvepoint\twelv@true\spaces@t}
\def\spaces@t{\relax%
 \iftwelv@ \ifsingl@\subspaces@t3:4;\else\subspaces@t1:1;\fi%
 \else \ifsingl@\subspaces@t3:5;\else\subspaces@t4:5;\fi \fi%
 \ifdoubl@ \multiply\baselineskip by 5%
 \divide\baselineskip by 4 \fi \unskip}
\def\subspaces@t#1:#2;{%
      \baselineskip = \normalbaselineskip%
      \multiply\baselineskip by #1 \divide\baselineskip by #2%
      \lineskip = \normallineskip%
      \multiply\lineskip by #1 \divide\lineskip by #2%
      \lineskiplimit = \normallineskiplimit%
      \multiply\lineskiplimit by #1 \divide\lineskiplimit by #2%
      \parskip = \normalparskip%
      \multiply\parskip by #1 \divide\parskip by #2%
      \abovedisplayskip = \normaldisplayskip%
      \multiply\abovedisplayskip by #1 \divide\abovedisplayskip by #2%
      \belowdisplayskip = \abovedisplayskip%
      \abovedisplayshortskip = \normaldispshortskip%
      \multiply\abovedisplayshortskip by #1%
	\divide\abovedisplayshortskip by #2%
      \belowdisplayshortskip = \abovedisplayshortskip%
      \advance\belowdisplayshortskip by \belowdisplayskip%
      \divide\belowdisplayshortskip by 2%
      \smallskipamount = \skipregister%
      \multiply\smallskipamount by #1 \divide\smallskipamount by #2%
      \medskipamount = \smallskipamount \multiply\medskipamount by 2%
      \bigskipamount = \smallskipamount \multiply\bigskipamount by 4 }
\def\normalbaselines{ \baselineskip=\normalbaselineskip%
   \lineskip=\normallineskip \lineskiplimit=\normallineskip%
   \iftwelv@\else \multiply\baselineskip by 4 \divide\baselineskip by 5%
     \multiply\lineskiplimit by 4 \divide\lineskiplimit by 5%
     \multiply\lineskip by 4 \divide\lineskip by 5 \fi }
\Twelvepoint  
\interlinepenalty=50
\interfootnotelinepenalty=5000
\predisplaypenalty=9000
\postdisplaypenalty=500
\hfuzz=1pt
\vfuzz=0.2pt
%
%
%
\def\pagecontents{%
   \ifvoid\topins\else\unvbox\topins\vskip\skip\topins\fi
   \dimen@ = \dp255 \unvbox255
   \ifvoid\footins\else\vskip\skip\footins\footrule\unvbox\footins\fi
   \ifr@ggedbottom \kern-\dimen@ \vfil \fi }
\def\makeheadline{\vbox to 0pt{ \skip@=\topskip
      \advance\skip@ by -12pt \advance\skip@ by -2\normalbaselineskip
      \vskip\skip@ \line{\vbox to 12pt{}\the\headline} \vss
      }\nointerlineskip}
\def\makefootline{\baselineskip = 1.5\normalbaselineskip
		 \line{\the\footline}}
\newif\iffrontpage
\newif\ifletterstyle
\newif\ifp@genum
\def\nopagenumbers{\p@genumfalse}
\def\pagenumbers{\p@genumtrue}
\pagenumbers
\newtoks\paperheadline
\newtoks\letterheadline
\newtoks\letterfrontheadline
\newtoks\lettermainheadline
\newtoks\paperfootline
\newtoks\letterfootline
\newtoks\date
\footline={\ifletterstyle\the\letterfootline\else\the\paperfootline\fi}
\paperfootline={\hss\iffrontpage\else\ifp@genum\tenrm
    -- \folio\ --\hss\fi\fi}
\letterfootline={\hfil}
\headline={\ifletterstyle\the\letterheadline\else\the\paperheadline\fi}
\paperheadline={\hfil}
\letterheadline{\iffrontpage\the\letterfrontheadline
     \else\the\lettermainheadline\fi}
\lettermainheadline={\rm\ifp@genum page \ \folio\fi\hfil\the\date}
\def\monthname{\relax\ifcase\month 0/\or January\or February\or
   March\or April\or May\or June\or July\or August\or September\or
   October\or November\or December\else\number\month/\fi}
\date={\monthname\ \number\day, \number\year}
\countdef\pagenumber=1  \pagenumber=1
\def\advancepageno{\global\advance\pageno by 1
   \ifnum\pagenumber<0 \global\advance\pagenumber by -1
    \else\global\advance\pagenumber by 1 \fi \global\frontpagefalse }
\def\folio{\ifnum\pagenumber<0 \romannumeral-\pagenumber
	   \else \number\pagenumber \fi }
\def\footrule{\dimen@=\prevdepth\nointerlineskip
   \vbox to 0pt{\vskip -0.25\baselineskip \hrule width 0.35\hsize \vss}
   \prevdepth=\dimen@ }
\newtoks\foottokens
\foottokens={\Tenpoint\singlespace}
\newdimen\footindent
\footindent=24pt
\def\vfootnote#1{\insert\footins\bgroup  \the\foottokens
   \interlinepenalty=\interfootnotelinepenalty \floatingpenalty=20000
   \splittopskip=\ht\strutbox \boxmaxdepth=\dp\strutbox
   \leftskip=\footindent \rightskip=\z@skip
   \parindent=0.5\footindent \parfillskip=0pt plus 1fil
   \spaceskip=\z@skip \xspaceskip=\z@skip
   \Textindent{$ #1 $}\footstrut\futurelet\next\fo@t}
\def\Textindent#1{\noindent\llap{#1\enspace}\ignorespaces}
\def\footnote#1{\attach{#1}\vfootnote{#1}}

\let\footsymbol=\star
\newcount\lastf@@t	     \lastf@@t=-1
\newcount\footsymbolcount    \footsymbolcount=0
\newif\ifPhysRev
\def\footsymbolgen{\relax \ifPhysRev \iffrontpage \NPsymbolgen\else
      \PRsymbolgen\fi \else \NPsymbolgen\fi
   \global\lastf@@t=\pageno \footsymbol }
\def\NPsymbolgen{\ifnum\footsymbolcount<0 \global\footsymbolcount=0\fi
   {\iffrontpage \else \advance\lastf@@t by 1 \fi
    \ifnum\lastf@@t<\pageno \global\footsymbolcount=0
     \else \global\advance\footsymbolcount by 1 \fi }
   \ifcase\footsymbolcount \fd@f\star\or \fd@f\dagger\or \fd@f\ast\or
    \fd@f\ddagger\or \fd@f\natural\or \fd@f\diamond\or \fd@f\bullet\or
    \fd@f\nabla\else \fd@f\dagger\global\footsymbolcount=0 \fi }
\def\fd@f#1{\xdef\footsymbol{#1}}
\def\PRsymbolgen{\ifnum\footsymbolcount>0 \global\footsymbolcount=0\fi
      \global\advance\footsymbolcount by -1
      \xdef\footsymbol{\sharp\number-\footsymbolcount} }
\def\space@ver#1{\let\@sf=\empty \ifmmode #1\else \ifhmode
   \edef\@sf{\spacefactor=\the\spacefactor}\unskip${}#1$\relax\fi\fi}
\def\attach#1{\space@ver{\strut^{\mkern 2mu #1} }\@sf\ }
%
%
%
\newcount\chapternumber	     \chapternumber=0
\newcount\sectionnumber	     \sectionnumber=0
\newcount\equanumber	     \equanumber=0
\let\chapterlabel=0
\newtoks\chapterstyle	     \chapterstyle={\Number}
\newskip\chapterskip	     \chapterskip=\bigskipamount
\newskip\sectionskip	     \sectionskip=\medskipamount
\newskip\headskip	     \headskip=8pt plus 3pt minus 3pt
\newdimen\chapterminspace    \chapterminspace=15pc
\newdimen\sectionminspace    \sectionminspace=10pc
\newdimen\referenceminspace  \referenceminspace=25pc
\def\chapterreset{\global\advance\chapternumber by 1
   \ifnum\equanumber<0 \else\global\equanumber=0\fi
   \sectionnumber=0 \makel@bel}
\def\makel@bel{\xdef\chapterlabel{%
\the\chapterstyle{\the\chapternumber}.}}
\def\sectionlabel{\number\sectionnumber \quad }
\def\alphabetic#1{\count255='140 \advance\count255 by #1\char\count255}
\def\Alphabetic#1{\count255='100 \advance\count255 by #1\char\count255}
\def\Roman#1{\uppercase\expandafter{\romannumeral #1}}
\def\roman#1{\romannumeral #1}
\def\Number#1{\number #1}
\def\unnumberedchapters{\let\makel@bel=\relax \let\chapterlabel=\relax
\let\sectionlabel=\relax \equanumber=-1 }
%
\def\titlestyle#1{\par\begingroup \interlinepenalty=9999
     \leftskip=0.03\hsize plus 0.20\hsize minus 0.03\hsize
     \rightskip=\leftskip \parfillskip=0pt
     \hyphenpenalty=9000 \exhyphenpenalty=9000
     \tolerance=9999 \pretolerance=9000
     \spaceskip=0.333em \xspaceskip=0.5em
     \iftwelv@\fourteenpoint\fourteenbf\else\twelvepoint\twelvebf\fi
     \noindent  #1\par\endgroup }

\def\spacecheck#1{\dimen@=\pagegoal\advance\dimen@ by -\pagetotal
   \ifdim\dimen@<#1 \ifdim\dimen@>0pt \vfil\break \fi\fi}
\def\chapter#1{\par \penalty-300 \vskip\chapterskip
   \spacecheck\chapterminspace
   \chapterreset \titlestyle{\chapterlabel \ #1}
   \nobreak\vskip\headskip \penalty 30000
   \wlog{\string\chapter\ \chapterlabel} }

%
\def\section#1{\par \ifnum\the\lastpenalty=30000\else
   \penalty-200\vskip\sectionskip \spacecheck\sectionminspace\fi
   \wlog{\string\section\ \chapterlabel \the\sectionnumber}
   \global\advance\sectionnumber by 1  \noindent
   {\caps\enspace\chapterlabel \sectionlabel #1}\par
   \nobreak\vskip\headskip \penalty 30000 }
\def\ssection#1{\par \ifnum\the\lastpenalty=30000\else
   \penalty-200\vskip\sectionskip \spacecheck\sectionminspace\fi
   \wlog{\string\section\ \chapterlabel \the\sectionnumber}
   \global\advance\sectionnumber by 1  \noindent
   {\S \caps\thinspace\chapterlabel \sectionlabel #1}\par
   \nobreak\vskip\headskip \penalty 30000 }
\def\subsection#1{\par
   \ifnum\the\lastpenalty=30000\else \penalty-100\smallskip \fi
   \noindent\undertext{#1}\enspace \vadjust{\penalty5000}}

\def\undertext#1{\vtop{\hbox{#1}\kern 1pt \hrule}}
\def\APPENDIX#1#2{\par\penalty-300\vskip\chapterskip
   \spacecheck\chapterminspace \chapterreset \xdef\chapterlabel{#1}
   \titlestyle{APPENDIX #2} \nobreak\vskip\headskip \penalty 30000
   \wlog{\string\Appendix\ \chapterlabel} }
\def\Appendix#1{\APPENDIX{#1}{#1}}
\def\appendix{\APPENDIX{A}{}}
%
%
%
\def\eqname#1{\relax \ifnum\equanumber<0
     \xdef#1{{\rm(\number-\equanumber)}}\global\advance\equanumber by -1
    \else \global\advance\equanumber by 1
      \xdef#1{{\rm(\chapterlabel \number\equanumber)}} \fi}
\def\eq{\eqname\?\?}
\def\eqn#1{\eqno\eqname{#1}#1}

\def\eqinsert#1{\noalign{\dimen@=\prevdepth \nointerlineskip
   \setbox0=\hbox to\displaywidth{\hfil #1}
   \vbox to 0pt{\vss\hbox{$\!\box0\!$}\kern-0.5\baselineskip}
   \prevdepth=\dimen@}}
\def\sequentialequations{\equanumber=-1}
%
%
\def\GENITEM#1;#2{\par \hangafter=0 \hangindent=#1
    \Textindent{#2}\ignorespaces}
\outer\def\newitem#1=#2;{\gdef#1{\GENITEM #2;}}
\newdimen\itemsize		  \itemsize=30pt
\newitem\item=1\itemsize;
\newitem\sitem=1.75\itemsize;	  
\newitem\ssitem=2.5\itemsize;	  
\outer\def\newlist#1=#2&#3&#4;{\toks0={#2}\toks1={#3}%
   \count255=\escapechar \escapechar=-1
   \alloc@0\list\countdef\insc@unt\listcount	 \listcount=0
   \edef#1{\par
      \countdef\listcount=\the\allocationnumber
      \advance\listcount by 1
      \hangafter=0 \hangindent=#4
      \Textindent{\the\toks0{\listcount}\the\toks1}}
   \expandafter\expandafter\expandafter
    \edef\c@t#1{begin}{\par
      \countdef\listcount=\the\allocationnumber \listcount=1
      \hangafter=0 \hangindent=#4
      \Textindent{\the\toks0{\listcount}\the\toks1}}
   \expandafter\expandafter\expandafter
    \edef\c@t#1{con}{\par \hangafter=0 \hangindent=#4 \noindent}
   \escapechar=\count255}
\def\c@t#1#2{\csname\string#1#2\endcsname}
\newlist\point=\Number&.&1.0\itemsize;
\newlist\subpoint=(\alphabetic&)&1.75\itemsize;
\newlist\subsubpoint=(\roman&)&2.5\itemsize;
%

%
%
%
\newcount\referencecount     \referencecount=0
\newif\ifreferenceopen	     \newwrite\referencewrite
\newtoks\rw@toks
\def\refmark#1{\relax\ifPhysRev\PRrefmark{#1}\else\NPrefmark{#1}\fi}
\def\refend{\refmark{\number\referencecount}}
\newcount\lastrefsbegincount \lastrefsbegincount=0
\def\refsend{\refmark{\count255=\referencecount
   \advance\count255 by-\lastrefsbegincount
   \ifcase\count255 \number\referencecount
   \or \number\lastrefsbegincount,\number\referencecount
   \else \number\lastrefsbegincount-\number\referencecount \fi}}
\def\refch@ck{\chardef\rw@write=\referencewrite
   \ifreferenceopen \else \referenceopentrue
   \immediate\openout\referencewrite=reference.aux \fi}
%
{\catcode`\^^M=\active 
  \gdef\obeyendofline{\catcode`\^^M\active \let^^M\ }}%
%
{\catcode`\^^M=\active 
  \gdef\ignoreendofline{\catcode`\^^M=5}}
{\obeyendofline\gdef\rw@start#1{\def\t@st{#1} \ifx\t@st\blankend%
\endgroup \@sf \relax \else \ifx\t@st\bl@nkend \endgroup \@sf \relax%
\else \rw@begin#1
\backtotext
\fi \fi } }
{\obeyendofline\gdef\rw@begin#1
{\def\n@xt{#1}\rw@toks={#1}\relax%
\rw@next}}
\def\blankend{}
{\obeylines\gdef\bl@nkend{
}}
\newif\iffirstrefline  \firstreflinetrue
\def\rwr@teswitch{\ifx\n@xt\blankend \let\n@xt=\rw@begin %
 \else\iffirstrefline \global\firstreflinefalse%
\immediate\write\rw@write{\noexpand\obeyendofline \the\rw@toks}%
\let\n@xt=\rw@begin%
      \else\ifx\n@xt\rw@@d \def\n@xt{\immediate\write\rw@write{%
	\noexpand\ignoreendofline}\endgroup \@sf}%
	     \else \immediate\write\rw@write{\the\rw@toks}%
	     \let\n@xt=\rw@begin\fi\fi \fi}
\def\rw@next{\rwr@teswitch\n@xt}
\def\rw@@d{\backtotext} \let\rw@end=\relax
\let\backtotext=\relax

\newdimen\refindent	\refindent=30pt
\def\refitem#1{\par \hangafter=0 \hangindent=\refindent \Textindent{#1}}
\def\REFNUM#1{\space@ver{}\refch@ck \firstreflinetrue%
 \global\advance\referencecount by 1 \xdef#1{\the\referencecount}}
\def\refnum#1{\space@ver{}\refch@ck \firstreflinetrue%
 \global\advance\referencecount by 1 \xdef#1{\the\referencecount}\refend}

\def\REF#1{\REFNUM#1%
 \immediate\write\referencewrite{%
 \noexpand\refitem{#1.}}%
\begingroup\obeyendofline\rw@start}
\def\ref{\refnum\?%
 \immediate\write\referencewrite{\noexpand\refitem{\?.}}%
\begingroup\obeyendofline\rw@start}
\def\Ref#1{\refnum#1%
 \immediate\write\referencewrite{\noexpand\refitem{#1.}}%
\begingroup\obeyendofline\rw@start}
\def\REFS#1{\REFNUM#1\global\lastrefsbegincount=\referencecount
\immediate\write\referencewrite{\noexpand\refitem{#1.}}%
\begingroup\obeyendofline\rw@start}
\newcount\figurecount	  \figurecount=0
\newif\iffigureopen	  \newwrite\figurewrite
\def\figch@ck{\chardef\rw@write=\figurewrite \iffigureopen\else
   \immediate\openout\figurewrite=figures.aux
   \figureopentrue\fi}
\def\FIGNUM#1{\space@ver{}\figch@ck \firstreflinetrue%
 \global\advance\figurecount by 1 \xdef#1{\the\figurecount}}
\def\FIG#1{\FIGNUM#1
   \immediate\write\figurewrite{\noexpand\refitem{#1.}}%
   \begingroup\obeyendofline\rw@start}
\def\fig{\FIGNUM\? fig.~\?%
\immediate\write\figurewrite{\noexpand\refitem{\?.}}%
\begingroup\obeyendofline\rw@start}
\def\figure{\FIGNUM\? figure~\?
   \immediate\write\figurewrite{\noexpand\refitem{\?.}}%
   \begingroup\obeyendofline\rw@start}
\def\Fig{\FIGNUM\? Fig.~\?%
\immediate\write\figurewrite{\noexpand\refitem{\?.}}%
\begingroup\obeyendofline\rw@start}
\def\Figure{\FIGNUM\? Figure~\?%
\immediate\write\figurewrite{\noexpand\refitem{\?.}}%
\begingroup\obeyendofline\rw@start}
\newcount\tablecount	 \tablecount=0
\newif\iftableopen	 \newwrite\tablewrite
\def\tabch@ck{\chardef\rw@write=\tablewrite \iftableopen\else
   \immediate\openout\tablewrite=tables.aux
   \tableopentrue\fi}
\def\TABNUM#1{\space@ver{}\tabch@ck \firstreflinetrue%
 \global\advance\tablecount by 1 \xdef#1{\the\tablecount}}
\def\TABLE#1{\TABNUM#1
   \immediate\write\tablewrite{\noexpand\refitem{#1.}}%
   \begingroup\obeyendofline\rw@start}
\def\Table{\TABNUM\? Table~\?%
\immediate\write\tablewrite{\noexpand\refitem{\?.}}%
\begingroup\obeyendofline\rw@start}
%

%
%
%
\def\masterreset{\global\pagenumber=1 \global\chapternumber=0
   \global\equanumber=0 \global\sectionnumber=0
   \global\referencecount=0 \global\figurecount=0 \global\tablecount=0 }
\def\FRONTPAGE{\ifvoid255\else\vfill\penalty-2000\fi
      \masterreset\global\frontpagetrue
      \global\lastf@@t=0 \global\footsymbolcount=0}

\def\paperstyle{\letterstylefalse\normalspace\papersize}
\def\letterstyle{\letterstyletrue\singlespace\lettersize}
\def\papersize{\hsize=35pc\vsize=50pc\hoffset=1pc\voffset=4pc
		\skip\footins=\bigskipamount}
\def\lettersize{\hsize=35.2pc\vsize=50.0pc\hoffset=0.5pc\voffset=2.5pc
   \skip\footins=\smallskipamount \multiply\skip\footins by 3 }
\paperstyle   
%
%
\def\MEMO{\letterstyle\FRONTPAGE \letterfrontheadline={\hfil}
    \line{\quad\fourteenrm KEK MEMORANDUM\hfil\twelverm\the\date\quad}
    \medskip \memod@f}

\def\memit@m#1{\smallskip \hangafter=0 \hangindent=1in
      \Textindent{\caps #1}}
\def\memod@f{\xdef\to{\memit@m{To:}}\xdef\from{\memit@m{From:}}%
     \xdef\topic{\memit@m{Topic:}}\xdef\subject{\memit@m{Subject:}}%
     \xdef\rule{\bigskip\hrule height 1pt\bigskip}}
\memod@f
%


%



%
\newskip\lettertopfil
\lettertopfil = 0pt plus 1.5in minus 0pt
\newskip\letterbottomfil
\letterbottomfil = 0pt plus 2.3in minus 0pt
\newskip\spskip \setbox0\hbox{\ } \spskip=-1\wd0
\def\addressee#1{\medskip\rightline{\the\date\hskip 30pt} \bigskip
   \vskip\lettertopfil
   \ialign to\hsize{\strut ##\hfil\tabskip 0pt plus \hsize \cr #1\crcr}
   \medskip\noindent\hskip\spskip}
\newskip\signatureskip	     \signatureskip=40pt
\def\signed#1{\par \penalty 9000 \bigskip \dt@pfalse
  \everycr={\noalign{\ifdt@p\vskip\signatureskip\global\dt@pfalse\fi}}
  \setbox0=\vbox{\singlespace \halign{\tabskip 0pt \strut ##\hfil\cr
   \noalign{\global\dt@ptrue}#1\crcr}}
  \line{\hskip 0.5\hsize minus 0.5\hsize \box0\hfil} \medskip }

\def\endletter{\ifnum\pagenumber=1 \vskip\letterbottomfil\supereject
\else \vfil\supereject \fi}
\newbox\letterb@x
\def\lettertext{\par\unvcopy\letterb@x\par}
\def\multiletter{\setbox\letterb@x=\vbox\bgroup
      \everypar{\vrule height 1\baselineskip depth 0pt width 0pt }
      \singlespace \topskip=\baselineskip }
\def\letterend{\par\egroup}
%
%
%
\newskip\frontpageskip
\newtoks\pubtype
\newtoks\Pubnum
\newtoks\pubnum
\newif\ifp@bblock \p@bblocktrue
\def\PH@SR@V{\doubl@true \baselineskip=24.1pt plus 0.2pt minus 0.1pt
	     \parskip= 3pt plus 2pt minus 1pt }
\def\PHYSREV{\paperstyle\PhysRevtrue\PH@SR@V}
\def\titlepage{\FRONTPAGE\paperstyle\ifPhysRev\PH@SR@V\fi
   \ifp@bblock\p@bblock\fi}
%
%
%
%
\def\nopubblock{\p@bblockfalse}
\def\endpage{\vfil\break}
\frontpageskip=1\medskipamount plus .5fil
\pubtype={ }
\newtoks\publevel
\publevel={Report}   
\Pubnum={\the\pubnum}
\pubnum={RRK-000}
%
%
\def\p@bblock{\begingroup \tabskip=\hsize minus \hsize
   \baselineskip=1.5\ht\strutbox \topspace-2\baselineskip
   \halign to\hsize{\strut ##\hfil\tabskip=0pt\crcr
   \the\Pubnum\cr \the\date\cr }\endgroup}

%
\def\title#1{\vskip\frontpageskip\vfill
   {\fourteenbf\titlestyle{#1}}\vskip\headskip\vfill }

\def\author#1{\vskip\frontpageskip\titlestyle{\twelvecp #1}\nobreak}

%
\def\address#1{\par\kern 5pt \titlestyle{\twelvepoint\sl #1}}
\def\andaddress{\par\kern 5pt \centerline{\sl and} \address}


%

%

%
%

%
\def\abstract#1{\vfill\vskip\frontpageskip\centerline{\fourteenrm ABSTRACT}
                \vskip\headskip#1\endpage}
%

%

%
%
%
\def\ie{\hbox{\it i.e.}}     
\def\eg{\hbox{\it e.g.}}     

\def\\{\relax\ifmmode\backslash\else$\backslash$\fi}
\def\globaleqnumbers{\relax\if\equanumber<0\else\global\equanumber=-1\fi}
\def\nextline{\unskip\nobreak\hskip\parfillskip\break}

\def\journal#1&#2(#3){\unskip, \sl #1~\bf #2 \rm (19#3) }

\def\topspace{\hrule height 0pt depth 0pt \vskip}

\let\int=\intop		
\def\prop{\mathrel{{\mathchoice{\pr@p\scriptstyle}{\pr@p\scriptstyle}{
		\pr@p\scriptscriptstyle}{\pr@p\scriptscriptstyle} }}}
\def\pr@p#1{\setbox0=\hbox{$\cal #1 \char'103$}
   \hbox{$\cal #1 \char'117$\kern-.4\wd0\box0}}
\def\lsim{\mathrel{\mathpalette\@versim<}}
\def\gsim{\mathrel{\mathpalette\@versim>}}
\def\@versim#1#2{\lower0.2ex\vbox{\baselineskip\z@skip\lineskip\z@skip
  \lineskiplimit\z@\ialign{$\m@th#1\hfil##\hfil$\crcr#2\crcr\sim\crcr}}}
%
%
%
\let\sec@nt=\sec
\def\sec{\relax\ifmmode\let\n@xt=\sec@nt\else\let\n@xt\section\fi\n@xt}
\def\obsolete#1{\message{Macro \string #1 is obsolete.}}
\def\firstsec#1{\obsolete\firstsec \section{#1}}
\def\firstsubsec#1{\obsolete\firstsubsec \subsection{#1}}
\def\thispage#1{\obsolete\thispage \global\pagenumber=#1\frontpagefalse}
\def\thischapter#1{\obsolete\thischapter \global\chapternumber=#1}
\def\nextequation#1{\obsolete\nextequation \global\equanumber=#1
   \ifnum\the\equanumber>0 \global\advance\equanumber by 1 \fi}
\def\BOXITEM{\afterassigment\B@XITEM\setbox0=}
\def\B@XITEM{\par\hangindent\wd0 \noindent\box0 }
%

%
\catcode`@=12 
\message{ by V.K.}


\relax

\def\yen{\hbox{Y\kern-0.75em =}}
%
%
\catcode`@=11
%
%
\font\fourteenmib=cmmib10 scaled\magstep2    \skewchar\fourteenmib='177
\font\twelvemib=cmmib10 scaled\magstep1	    \skewchar\twelvemib='177
\font\elevenmib=cmmib10 scaled\magstephalf   \skewchar\elevenmib='177
\font\tenmib=cmmib10			    \skewchar\tenmib='177
%
\font\fourteenbsy=cmbsy10 scaled\magstep2     \skewchar\fourteenbsy='60
\font\twelvebsy=cmbsy10 scaled\magstep1	      \skewchar\twelvebsy='60
\font\elevenbsy=cmbsy10 scaled\magstephalf    \skewchar\elevenbsy='60
\font\tenbsy=cmbsy10			      \skewchar\tenbsy='60
%
\newfam\mibfam
\def\samef@nt{\relax \ifcase\f@ntkey \rm \or\oldstyle \or\or
	 \or\it \or\sl \or\bf \or\tt \or\caps \or\mib \fi }
\def\fourteenpoint{\relax
    \textfont0=\fourteenrm	    \scriptfont0=\tenrm
    \scriptscriptfont0=\sevenrm
     \def\rm{\fam0 \fourteenrm \f@ntkey=0 }\relax
    \textfont1=\fourteeni	    \scriptfont1=\teni
    \scriptscriptfont1=\seveni
     \def\oldstyle{\fam1 \fourteeni\f@ntkey=1 }\relax
    \textfont2=\fourteensy	    \scriptfont2=\tensy
    \scriptscriptfont2=\sevensy
    \textfont3=\fourteenex     \scriptfont3=\fourteenex
    \scriptscriptfont3=\fourteenex
    \def\it{\fam\itfam \fourteenit\f@ntkey=4 }\textfont\itfam=\fourteenit
    \def\sl{\fam\slfam \fourteensl\f@ntkey=5 }\textfont\slfam=\fourteensl
    \scriptfont\slfam=\tensl
    \def\bf{\fam\bffam \fourteenbf\f@ntkey=6 }\textfont\bffam=\fourteenbf
    \scriptfont\bffam=\tenbf	 \scriptscriptfont\bffam=\sevenbf
    \def\tt{\fam\ttfam \twelvett \f@ntkey=7 }\textfont\ttfam=\twelvett
    \h@big=11.9\p@{} \h@Big=16.1\p@{} \h@bigg=20.3\p@{} \h@Bigg=24.5\p@{}
    \def\caps{\fam\cpfam \twelvecp \f@ntkey=8 }\textfont\cpfam=\twelvecp
    \setbox\strutbox=\hbox{\vrule height 12pt depth 5pt width\z@}
    \def\mib{\fam\mibfam \fourteenmib \f@ntkey=9 }
    \textfont\mibfam=\fourteenmib      \scriptfont\mibfam=\tenmib
    \scriptscriptfont\mibfam=\tenmib
    \samef@nt}
\def\twelvepoint{\relax
    \textfont0=\twelverm	  \scriptfont0=\ninerm
    \scriptscriptfont0=\sixrm
     \def\rm{\fam0 \twelverm \f@ntkey=0 }\relax
    \textfont1=\twelvei		  \scriptfont1=\ninei
    \scriptscriptfont1=\sixi
     \def\oldstyle{\fam1 \twelvei\f@ntkey=1 }\relax
    \textfont2=\twelvesy	  \scriptfont2=\ninesy
    \scriptscriptfont2=\sixsy
    \textfont3=\twelveex	  \scriptfont3=\twelveex
    \scriptscriptfont3=\twelveex
    \def\it{\fam\itfam \twelveit \f@ntkey=4 }\textfont\itfam=\twelveit
    \def\sl{\fam\slfam \twelvesl \f@ntkey=5 }\textfont\slfam=\twelvesl
    \scriptfont\slfam=\ninesl
    \def\bf{\fam\bffam \twelvebf \f@ntkey=6 }\textfont\bffam=\twelvebf
    \scriptfont\bffam=\ninebf	  \scriptscriptfont\bffam=\sixbf
    \def\tt{\fam\ttfam \twelvett \f@ntkey=7 }\textfont\ttfam=\twelvett
    \h@big=10.2\p@{}
    \h@Big=13.8\p@{}
    \h@bigg=17.4\p@{}
    \h@Bigg=21.0\p@{}
    \def\caps{\fam\cpfam \twelvecp \f@ntkey=8 }\textfont\cpfam=\twelvecp
    \setbox\strutbox=\hbox{\vrule height 10pt depth 4pt width\z@}
    \def\mib{\fam\mibfam \twelvemib \f@ntkey=9 }
    \textfont\mibfam=\twelvemib	    \scriptfont\mibfam=\tenmib
    \scriptscriptfont\mibfam=\tenmib
    \samef@nt}
\def\tenpoint{\relax
    \textfont0=\tenrm	       \scriptfont0=\sevenrm
    \scriptscriptfont0=\fiverm
    \def\rm{\fam0 \tenrm \f@ntkey=0 }\relax
    \textfont1=\teni	       \scriptfont1=\seveni
    \scriptscriptfont1=\fivei
    \def\oldstyle{\fam1 \teni \f@ntkey=1 }\relax
    \textfont2=\tensy	       \scriptfont2=\sevensy
    \scriptscriptfont2=\fivesy
    \textfont3=\tenex	       \scriptfont3=\tenex
    \scriptscriptfont3=\tenex
    \def\it{\fam\itfam \tenit \f@ntkey=4 }\textfont\itfam=\tenit
    \def\sl{\fam\slfam \tensl \f@ntkey=5 }\textfont\slfam=\tensl
    \def\bf{\fam\bffam \tenbf \f@ntkey=6 }\textfont\bffam=\tenbf
    \scriptfont\bffam=\sevenbf	   \scriptscriptfont\bffam=\fivebf
    \def\tt{\fam\ttfam \tentt \f@ntkey=7 }\textfont\ttfam=\tentt
    \def\caps{\fam\cpfam \tencp \f@ntkey=8 }\textfont\cpfam=\tencp
    \setbox\strutbox=\hbox{\vrule height 8.5pt depth 3.5pt width\z@}
    \def\mib{\fam\mibfam \tenmib \f@ntkey=9 }
    \textfont\mibfam=\tenmib   \scriptfont\mibfam=\tenmib
    \scriptscriptfont\mibfam=\tenmib
    \samef@nt}
%
%
\Twelvepoint
\catcode`@=12
%
\everyjob{\input ritpdef\message{Good Luck}}
\message{ modified by K-I. A.}



\def\B#1{\Roman #1}
\def\BA#1#2{\Roman #1$\,(#2)$}
\def\proclami #1#2#3\par{\medbreak
  \nobreak\noindent{\bf#1\enspace}{#2\enspace}{\sl#3}\par\medbreak}
\def\rn#1{\expandafter{\romannumeral #1}}


\def\refmark#1{{[#1]}}


\REF\RS{M.~P.~Jr.~Ryan and L.~C.~Shepley, {\sl Homogeneous
Relativistic Cosmologies}, (Princeton University Press, Princeton,
1975).}
\REF\KSMH{D.~Kramer, H.~Stephani, M.~MacCallum and E.~Herlt, {\sl
Exact Solutions of Einstein's Field Equations}, (Cambridge University
Press, Cambridge, 1980).}
\REF\TL{W.~P.~Thurston, {\sl The Geometry and Topology
of 3-manifolds}, to be published by Princeton University Press,
1978/79.}
\REF\TP{W.~P.~Thurston, {\sl Bull.~Amer.~Math.~Soc.~\bf 6}
(1982) 357.}
\REF\S{P.~Scott, {\sl Bull.~London Math.~Soc.~\bf 15}
(1983) 401.}
\REF\M{J.~Milnor, {\sl Adv.~in Math.~\bf 21} (1976) 293.}
\REF\F{H.~V.~Fagundes, {\sl Phys.~Rev.~Lett.~\bf 54}
(1985) 1200; \hfil\break
{\sl Gen.~Rel.~Grav.~\bf 24} (1992) 199.}
\REF\AS{A.~Ashtekar and J.~Samuel, {\sl Class.~Quantum~Grav.~\bf 8}
(1991) 2191.}
\REF\MT{M.~A.~H.~MacCallum and A.~H.~Taub, {\sl
Commun.~Math.~Phys.~\bf 25} (1972) 173.}
\REF\M{V.~Moncrief, {\sl J.~Math.~Phys.~\bf 30} (1989) 2907.}
\REF\HN{A.~Hosoya and K.~Nakao,
{\sl Class.~Quantum~Grav.~\bf 7}  (1990) 163; \hfil\break
{\sl Prog.~Theor.~Phys.~\bf 84} (1990) 739, \hfil\break
Y.~Fujiwara and J.~Soda, {\sl Prog.~Theor.~Phys.~\bf 83} (1990) 733.}
\REF\IO{T.~Okamura and H.~Ishihara, {\sl Phys.~Rev.~\bf D46} (1992)
572; to appear in {\sl Phys.~Rev.}}
\REF\HKT{A.~Hosoya, T.~Koike and M.~Tanimoto, private
communication.}



\publevel={preprint}
\pubnum={KUCP-55}

\titlepage
\title{Comments on Closed Bianchi Models}
\author{Yoshihisa FUJIWARA, Hideki ISHIHARA, Hideo KODAMA}
\address{Department of Fundamental Sciences, FIHS,\nextline
Kyoto University, Kyoto, Japan}
\abstract{We show several kinematical properties that are intrinsic
to the Bianchi models with compact spatial sections. Especially, with
spacelike hypersurfaces being closed, (A)~no anisotropic
expansion is allowed for Bianchi type \B 5 and \BA 7{A\not=0}, and
(B)~type \B 4 and \BA 6{A\not=0,1} does not exist. In order to show
them, we put into geometric terms what is meant by spatial homogeneity
and employ a mathematical result on 3-manifolds. We make clear the
relation between the Bianchi type symmetry of space-time and spatial
compactness, some part of which seem to be unnoticed in the
literature. Especially, it is shown under what conditions class~B
Bianchi models do not possess compact spatial sections. Finally we
briefly describe how this study is useful in investigating global
dynamics in (3+1)-dimensional gravity.}

\sequentialequations


The so-called Bianchi model in general relativity has been widely
studied by many people (see \refmark\RS \refmark\KSMH\ for
example). In the model, a space-time is assumed to have symmetry,
namely spatially homogeneity. Since the gravitational degrees of
freedom are reduced to be finite in this model, it gives an
appropriate system for us to understand the full complexity of the
Einstein equations and physical cosmology described by it. It also
serves as a natural background for perturbations and for a study of
quantum fields on it. Moreover, this minisuperspace can be used as a
toy model for understanding quantum cosmology. In that context one
usually assumes that the spatially homogeneous hypersurface is compact
without boundary, \ie~closed.

In this paper, we study such a closed Bianchi model from geometrical
viewpoint. We show some interesting kinematical properties intrinsic
to the closed Bianchi model which follow from the spatial
homogeneity and spatial compactness. In order to show them, we utilize
a recent mathematical result on three-dimensional manifold by Thurston
and others. Some of our results were mentioned in some works by other
people, but under somewhat restrictive conditions. We do not assume
such restrictions in our geometrical argument and shall make clear
some points unnoticed in the literature. In the end of our
paper, we describe how such study of closed Bianchi models is useful
in order to investigate the global dynamical degrees of freedom, which
are often called \lq\lq moduli\rq\rq, of a closed 3-space in
(3+1)-dimensional gravity.

Let us begin with a general description of homogeneity in geometry. A
Riemannian manifold $(M,g)$ is defined to be {\sl locally
homogeneous}\/, if for every pair of points $x,y\in M$ there are
neighborhoods $U$ and $V$ of $x$ and $y$, for which there exists a
local isometry mapping $(U,x)$ to $(V,y)$. Such local isometries do
not, in general, extend to isometries of the whole $(M,g)$. Let us
denote the {\sl full} group of isometries of $(M,g)$ by
$\hbox{Isom}(M,g)$. Then if for every pair of points $x,y\in M$ there
is an isometry $\Phi\in\hbox{Isom}(M,g)$ such that $\Phi(x)=y$, that
is, $\hbox{Isom}(M,g)$ acts transitively on $M$, one says that $(M,g)$
is {\sl (globally) homogeneous}\/. Since it is a standard fact that a
simply-connected and locally homogeneous manifold is globally
homogeneous, the universal covering space $(\widetilde M,\widetilde
g)$ of a locally homogeneous manifold $(M,g)$ must be homogeneous.
For our convenience, we shall say that a homogeneous manifold $(M,g)$
is {\sl simply homogeneous} if $\hbox{Isom}(M,g)$ has a
three-dimensional subgroup $G_3$ that acts {\sl simply-transitively}
on $M$. In that case, we call the $G_3$ a {\sl homogeneity group}\/.
If the universal covering space of a locally homogeneous manifold is
simply homogeneous with a homogeneity group $G_3$, then we shall call
it a locally homogeneous manifold with a homogeneity group $G_3$.

The usual Bianchi model starts with the following assumptions for
spatial homogeneity. A three-dimensional Lie group $G_3$ acts on a
space-time as a group of isometries of the space-time, such that each
orbit is a spacelike hypersurface $\Sigma$ on which $G_3$ acts {\sl
simply-transitively}\/. The space-time considered is topologically a
product space $\Sigma\times{\bf R}$. We have a family of 3-manifolds
$(\Sigma,g(t))$ for each $t\in{\bf R}$, where $g$ is the spatial
metric intrinsic to $\Sigma$. Then it follows that $G_3$ acts
simply-transitively on each $(\Sigma,g(t))$ as a group of isometries
of $(\Sigma,g(t))$. In the above terms, $(\Sigma,g(t))$ is simply
homogeneous.

Such a Riemannian manifold is well understood in the context of
Bianchi cosmology. (See \refmark\RS \refmark\KSMH\ for
references and complete information on the definitions and the
notations below.) Denote by $C^K_{IJ}\ (I,J,K=1\sim 3)$ the structure
constant of the Lie algebra of $G_3$ with respect to a certain basis
$\{\xi_I\}$: $[\xi_I,\xi_J]=C^K_{IJ}\,\xi_K$. Then there {\sl
globally} exists on $M$ an invariant basis $\{X_I\}$ which one can
choose so that
     $$
     [X_I,X_J]= -C^K_{IJ}\,X_K,
     \eqn\comrelation
     $$
and its invariant dual basis $\{\chi^I\}$ which satisfies the
Mauer-Cartan equation
     $$
     d\chi^I= {1\over 2}\, C^I_{JK}\,\chi^J\wedge\chi^K.
     \eqno\eq
     $$
The metric $g$ on $M$ can be expressed in terms of $\chi^I$ as
     $$
     ds^2=g_{IJ}\,\chi^I\,\chi^J,
     \eqn\metric
     $$
where $g_{IJ}$ is a nonsingular constant matrix. Thus the Riemannian
metric of a simply homogeneous manifold can be completely specified
by the Lie algebra of $G_3$ and a constant matrix $g_{IJ}$.

The three-dimensional real Lie algebras are completely classified into
the well-known Bianchi types. They are denoted as \B 1, \B 2, \B 3,
\B 4, \B 5, \BA 6A, \BA 7A, \B 8 and \B 9. Here \BA 6A and \BA 7A are
one-parameter families of algebras and \B 3 is isomorphic to
\BA 6{A=1}.  They are subdivided into class~A and~B according to
whether the trace of the structure constant $a_I\equiv{1\over 2}\,
C^J_{IJ}$ has a vanishing norm $(\delta^{IJ}\,a_I a_J)^{1/2}$ or not.
Class~A consists of  \B 1, \B 2, \BA 60, \BA 70, \B 8 and \B 9,
while the other types, \B 4, \B 5, \B 3=\BA 61,
\BA 6{A\not=0}, \BA 7{A\not=0} belong to class~B.

For a given Bianchi Lie algebras \B 1$\sim$\B 9, there exists a unique
(up to a constant matrix $g_{IJ}$) {\sl simply-connected} Riemannian
manifold $M$ diffeomorphic to a simply-connected Lie group $G$ which
is uniquely determined by the given algebra. $G$ is then a group of
isometries acting simply-transitively on $M$. Therefore, for a
simply-connected and simply homogeneous manifold with a homogeneity
group $G_3$, its topology is determined by the Bianchi type of the
Lie algebra of $G_3$ while its metric is given by \metric. Especially,
$M$ is diffeomorphic to either ${\bf R}^3$ or $S^3$ depending
on whether $G_3$ is of the type \B 1$\sim$\B 8 or \B 9 respectively.

However, it is not adequate for a general study of spatially
homogeneous space-times to restrict only on those simply homogeneous
manifolds as spatial homogeneous sections. In fact, when one can
construct a compact manifold by identifying certain points in a simply
homogeneous manifold, it usually lowers the dimension of the group of
isometries so that the resulting manifold is not simply homogeneous
any longer. For example, a homogeneous spatial section in every
class~B Bianchi model cannot be simply homogeneous if it is compact,
as we shall see later. Therefore we should also include in our
consideration a locally homogeneous manifold whose universal covering
space is simply homogeneous. More explicitly, we shall consider a
wider class of spatially homogeneous space-times as follows. A
three-dimensional Lie group $G_3$ is now assumed to act on the {\sl
universal covering space} of a space-time as a group of isometries,
such that each orbit is a spacelike hypersurface $\widetilde\Sigma$ on
which $G_3$ acts simply-transitively. Then $G_3$ acts
simply-transitively on each $\widetilde\Sigma$ as a group of
isometries of $\widetilde\Sigma$, so the underlying manifold $\Sigma$
is a locally homogeneous manifold with a homogeneity group $G_3$. We
will concentrate on the study of $\Sigma$ in the following, which is
denoted by $(M,g)$. Henceforth we assume that $M$ is closed.

Let us now study in general a locally homogeneous 3-manifold $(M,g)$
with a homogeneity group $G_3$. Consider its universal
covering space $(\widetilde M,\widetilde g)$ with a covering map $p$.
A covering transformation is a homeomorphism $\gamma\colon{\widetilde
M}\rightarrow\widetilde M$ such that $p\circ\gamma=p$. The set of
covering transformations is a group under composition, which is
called a covering transformation group $\Gamma$. Because in this
case $\Gamma$ is a discrete subgroup of $\hbox{Isom}(\widetilde
M,\widetilde g)$, acting freely and properly discontinuously on
$\widetilde M$, $M$ is isometric to the quotient space $\widetilde
M/\Gamma$. Thus in order to study $M$ one has to examine $\Gamma$ and
$\widetilde M$.  $\widetilde M$ is, as shown above, simply homogeneous
so that its structure is determined by the homogeneity group $G_3$.
Note that $G_3$ is in general a subgroup of $\hbox{Isom} (\widetilde
M,\widetilde g)$. In what follows, we will classify all the possible
geometries of $(\widetilde M,\widetilde g)$ by employing a modern
viewpoint of \lq\lq geometry\rq\rq\ and show some relations between
those geometries and the Bianchi types. How one can choose $\Gamma$ to
obtain a compact locally homogeneous manifold $M\cong{\widetilde
M}/\Gamma$ depends on each class of geometry.

In a modern approach, \lq\lq geometry\rq\rq\ can be viewed in the
following way (see \refmark\TL \refmark\TP \refmark\S, for
example). Suppose that $X$ is a manifold and $G$ is a group acting on
$X$. $G$ is assumed to act transitively on $X$ with compact point
stabilizer (the isotropy subgroup of $G$ at any point of $X$ is
compact). Then a \lq\lq geometry\rq\rq\ is the pair $(X,G)$ and the
properties of $X$ invariant under the action of $G$. One can recover
the ordinary viewpoint of differential geometry by finding a
$G$-invariant metric on $X$. (Its existence is guaranteed since $G$'s
stabilizer is compact at every point. And as $G$ acts transitively on
$X$, the metric is complete.) In general, there would be many
different $G$-invariant metrics on $X$ so that $X$ can have a variety
of properties. For instance, as is easily seen from above, the
universal covering $(\widetilde M,\widetilde g)$ of a locally
homogeneous $(M,g)$ and its isometry group $\hbox{Isom}(\widetilde
M,\widetilde g)$ are an example of $(X,G)$.

Thurston \refmark\TP\ classified all the three-dimensional
geometries under the following restrictions. In this classification,
two geometries $(X,G)$ and $(X',G')$ are defined to be equivalent if
there is a diffeomorphism of $X$ with $X'$ which casts the action of
$G$ on $X$ onto that of $G'$ on $X'$. First of all,
$X$ is assumed simply-connected for one can always study universal
covering spaces if necessary. Secondly, we restrict ourselves to the
case that $G$ is maximal. It means that when two groups $G_1$ and
$G_2$ such that $G_1\subset G_2$ can both act on $X$, one should take
$G_2$ as $G$. In our consideration above, $G=\hbox{Isom}(\widetilde
M,\widetilde g)$ is maximal when one chooses $\widetilde g$
appropriately by changing $g_{IJ}$ in the metric
\metric\ so that the full isometry group $\hbox{Isom}(\widetilde
M,\widetilde g)$ becomes maximum. So, for example,
$(E^3,{\bf R}^3)$ is out of consideration, where $E^3$ is the
Euclidean space and ${\bf R}^3$ acts on it as translations. Rather,
one should take the full isometry group of $E^3$, namely the
three-dimensional Euclidean group $E(3)$, as $G$.  Finally, it is
assumed that $G$ has a subgroup $\Gamma$ which acts on $X$ as a
covering group so that $X/\Gamma$ becomes compact. Then the geometry
is said to admit a compact quotient. Now the Thurston's theorem can be
stated as follows:

\proclami {Theorem}{(Thurston)} Any maximal and simply-connected
three-dimensional geometry which admits a compact quotient is
equivalent to one of the eight geometries $(X,\hbox{Isom}X)$ described
below. \par
\item{(\rn 1)} $X=S^3$, the spherical geometry.
$\hbox{Isom}X=SO(4)$.
\item{(\rn 2)} $X=E^3$, the Euclidean geometry.
$\hbox{Isom}X=E(3)^+$.
\item{(\rn 3)} $X=H^3$, the hyperbolic geometry.
$\hbox{Isom}X=PSL(2,{\bf C})$.
\item{(\rn 4)} $X=S^2\times E^1$. $\hbox{Isom}X=(\hbox{Isom}S^2\times
\hbox{Isom}E^1)^+$.
\item{(\rn 5)} $X=H^2\times E^1$. $\hbox{Isom}X=(\hbox{Isom}H^2\times
\hbox{Isom}E^1)^+$.
\item{(\rn 6)} $X=\widetilde{T_1(H^2)}$, the universal covering space
of unit tangent space of $H^2$.
$\hbox{Isom}X=\widetilde{\hbox{Isom}H^2}\times{\bf R}$.
\item{(\rn 7)} $X=Nil$, the Heisenberg group.
\item{(\rn 8)} $X=Sol$, a solvable three-dimensional Lie group.

Some remarks follow. Here $\hbox{Isom}X$ means only orientation
preserving isometries. $\Gamma$ should be a discrete subgroup of this
$\hbox{Isom}X$ in order that $X/\Gamma$ be orientable. The isometries
corresponding to (\rn 7) and (\rn 8) are omitted. The dimension of
$\hbox{Isom}X$ is 6 for (\rn 1)$\sim$(\rn 3), 4 for (\rn 4)$\sim$(\rn
7), and 3 for (\rn 8). $Nil$ and $Sol$ spaces are three-dimensional
Lie groups whose Lie algebra is of Bianchi type \B 2 and \BA 60
respectively. $\widetilde{T_1(H^2)}$ is equivalent to
$\widetilde{SL_2{\bf R}}$ which can be also regarded as the universal
covering of the Lie group of Bianchi type \B 8. Final remark is that
all the possibilities of a compact quotients are known for all the
classes except the hyperbolic geometry (\rn 3). For complete
information, the reader should refer to \refmark\TP \refmark\S.

Returning to our study of locally homogeneous manifolds, we recall
that the homogeneity group $G_3$ is assumed to be a subgroup of
$\hbox{Isom}(\widetilde M,\widetilde g)$ for any metric of the form
\metric. Since in the Thurston's classification
$G=\hbox{Isom}(\widetilde M,\widetilde g)$ must be maximal, $G_3$ is a
subgroup of $G$ which belongs to one of the eight classes. From this
we can deduce two interesting consequences.

\item{(A)}
No anisotropic expansion is allowed for Bianchi model \B 5
with a closed spatial section.

\noindent
Consider the universal covering space $(\widetilde M,\widetilde g)$ of
a closed spatial section that is a locally homogeneous spacelike
hypersurface in Bianchi \B 5 space-time. Since $(\widetilde
M,\widetilde g)$ is simply homogeneous, the metric $\widetilde
g$ is of the form \metric. According to Milnor \refmark\M(Special
Example 1.7), for  the Lie algebra of Bianchi type \B 5, $(\widetilde
M,\widetilde g)$ is necessarily isometric to a maximally symmetric
space with a negative constant curvature of a certain magnitude,
whatever one chooses as $g_{IJ}$ in \metric. Equivalently,
$(\widetilde M,\widetilde g)$ is always the hyperbolic geometry (\rn
3) in the above classification. This fact means that each
locally homogeneous spacelike hypersurface in Bianchi \B 5 space-time
is locally isometric to a negative constant curvature space. But it
does not immediately imply that anisotropic expansion is impossible
(\eg\ it does not for Bianchi \B 1 model whose homogeneous spatial
section is flat). For there possibly remains a continuous choice of
the covering transformation group $\Gamma$ for constructing
$M\cong{\widetilde M}/\Gamma$. In other words, we may have the freedom
of \lq\lq moduli\rq\rq\ of a closed manifold. However, no such freedom
arises in this hyperbolic geometry due to the Mostow rigidity theorem
(see \refmark\TL \refmark\TP). The theorem asserts that if two
closed manifolds with hyperbolic geometry are homeomorphic to each
other, they are actually isometric to each other. So a Bianchi \B 5
homogeneous universe is rigid allowing only a change of overall scale
factor.  Thus the proposition was proved.

\medskip
\item{(B)}
Bianchi \B 4 and \BA 6{A\not=0,1} with a closed spatial section does
not exist.

\noindent
The proof is easy. The homogeneity group must be a subgroup of one of
the eight isometry groups in the classification. However, it cannot be
for the homogeneity group of Bianchi type \B 4 or \BA 6{A\not=0,1}, as
is explicitly shown by examining the Lie algebras of them and the
above eight classes. Therefore, no closed Bianchi \B 4 and \BA
6{A\not=0,1} exists.

\medskip
We can proceed further and study the correspondence between the
Bianchi models and the Thurston's eight geometries. The result is
summarized in Table~1 and Table~2. This can be obtained from the
consideration of the proof of the Thurston's classification theorem
\refmark\S. From the table it can be seen that a closed locally
homogeneous manifold with the homogeneity group \BA 7{A\not=0} admits
the hyperbolic geometry as the case of \B 5 so that no anisotropic
expansion is admitted also for this type \BA 7{A\not=0}. Also note
that the geometry of (\rn 4) $S^2\times E^1$ corresponds to the
Kantowski-Sachs model, rather than the Bianchi models considered here.

Essentially the same correspondence between the Thurston's
classification and the Bianchi types was presented by Fagundes
\refmark\F. The fact (B) was also mentioned in it. However, it
was assumed that the covering transformation group $\Gamma$ is always
a discrete subgroup of the homogeneity group $G_3$. (This is not true
in general in our assumptions, and causes a substantial difference for
class~B models as is explained below.)  And our approach here is a
group theoretical one which clarifies the relation among $G_3$,
$\Gamma$ and the eight maximal groups in the Thurston's
classification.  Such an approach is also useful in studying the
compactification problem of class~B models, as we shall see in the
sequel.

On the other hand, in a more general study of Bianchi cosmology with
compact spatial sections, Ashtekar and Samuel \refmark\AS\ showed
that class~B Bianchi universe cannot be closed under a certain
condition. We first describe the condition and then put it into a group
theoretical expression. By doing it, we can examine the possibility
for the compactification of class~B models more closely.

On a locally homogeneous manifold $(M,g)$, in general, one can define
an invariant basis only locally on each patch of its open coverings
$\{U_i\}$. Because each patch $U_i$ is isometric to an open
neighborhood $\widetilde U_i$ in $\widetilde M$ where an invariant
basis can be globally defined, one can induce an invariant basis on
each $U_i$. Denote by $X_I$ and $X'_I$ such two induced invariant
basis on $U_i$ and $U_j$. Now, at every point in the overlapping
region $U_i\cap U_j\not=\phi$, suppose that these two basis are
related with each other as $X'_I=X_J\, T^J{}_I$ where $T^I{}_J$ is a
constant matrix. In other words, the transition between the two
invariant bases is a Lie algebra automorphism. That this condition
holds for every overlaps of $\{U_i\}$ is the additional condition
mentioned above.  As was proved in \refmark\AS, any class~B models
cannot be compactified provided that this condition holds.

The above condition of Lie algebra automorphism can be put into a
group theoretical expression if one examines the universal covering
space $(\widetilde M,\widetilde g)$ and its covering transformation
group $\Gamma$. That is, {\sl $\Gamma$ is a subgroup of the normalizer
of $G_3$ in $\hbox{Isom}(\widetilde M,\widetilde g)$},
or equivalently,
     $$
     \Gamma\subset N(G_3),
     \eqn\condition
     $$
where $N(G_3)\equiv\{
g\in\hbox{Isom}(\widetilde M,\widetilde g)| g\,G_3\,g^{-1}=G_3 \}$.
This can be easily seen as follows. Let us pay attention to the
identification by the action of $\Gamma$ on $\widetilde M$. For
definiteness, we assume that $G_3$ or $\Gamma$ acts on $\widetilde M$
as right transformation $R_h\colon x\rightarrow xh$, ($x\in\widetilde
M$, $h\in G_3,\,\Gamma$). Fix an element $\gamma\in\Gamma$. An
invariant basis $X_I$ can be globally expanded on $\widetilde M$ by
choosing a point $y\in\widetilde M$ and setting
$X_{I\,x}\equiv(R_g)_*\,X_{I\,y}$ at every point $x\in\widetilde M$,
where $g$ is a unique element of $G_3$ such that $x=yg$. Now
take the point $y\in\widetilde M$ and the identified point
$y\gamma\in\widetilde M$, then
     $$
     (R_{\gamma^{-1}})_*\, X_{I\,y\gamma}=T^J{}_I\, X_{J\,y}.
     \eqno\eq
     $$
Here we have just written the induced vector on the left-hand side as
a linear combination of the basis $X_I$ at $y$. In the next
place, take $x$ and $x\gamma$. Because $G_3$ acts simply-transitively
on $\widetilde M$, there exists a unique element $g\in G_3$ such that
$x\gamma=(y\gamma) g$. Now supposing that $\Gamma\subset N(G_3)$,
there exists an element $g'\in G_3$ such that $\gamma g=g'\gamma$. One
has then (a) $x\gamma=yg'\gamma$ so that $x=yg'$ since $\gamma$ is a
covering transformation, and (b) $R_{\gamma^{-1}}\,
R_g=R_{g\gamma^{-1}}=R_{\gamma^{-1}g'}= R_{g'}\, R_{\gamma^{-1}}$.
{}From these (a) and (b), it follows that %
     $$
     \eqalign{
     (R_{\gamma^{-1}})_* & \, X_{I\,x\gamma}=
   	  (R_{g'})_*\, (R_{\gamma^{-1}})_*\, X_{I\,y\gamma} \cr
     & = (R_{g'})_*\, X_{J\,y}\, T^J{}_I  \cr
     & = X_{J\,x}\, T^J{}_I,  \cr
     }\eqn\Xtrans
     $$
which means that the invariant basis at a point and the induced basis
from the identified point are related with each other as a Lie algebra
automorphism. This holds for every element $\gamma\in\Gamma$, if
$\Gamma\subset N(G_3)$.

{}From this group theoretical viewpoint, we would like to understand why
class~B models cannot be compactified under this condition. Again
consider the universal covering  $(\widetilde M,\widetilde g)$. First
of all, we note that the next equality holds:
     $$
     {\cal L}_{X_I}\Omega=C_I\, \Omega,
     \eqn\LiedOmega
     $$
where $\Omega$ is the volume element
$(1/3!)\, \epsilon_{IJK}\, \chi^I\wedge\chi^J\wedge\chi^K$ and
$C_I\equiv C^J_{IJ}$, the trace of the structure constant. Since
$X_I$ is globally defined on $\widetilde M$, \LiedOmega\ means that
$X_I$ generates a 1-parameter group of diffeomorphisms of $\widetilde
M$ onto itself, which changes the volume. Remember that $C_I$ is not
zero for class~B types. Now we show that when \condition\ holds, a
vector field which is defined by $V\equiv X_I C_J\,g^{IJ}\equiv X_I
C^I$ globally exists on $M$ and it satisfies
     $$
     {\cal L}_V\, \Omega=(C_I C^I)\,\Omega,
     \eqn\VLiedOmega
     $$
This \VLiedOmega\ follows immediately from \LiedOmega\ because
\LiedOmega\ locally holds so does its projection onto $M$. The global
existence of $V$ on $M$ can be shown as follows. The covering
transformation $\gamma\in\Gamma$ generates a diffeomorphism
$\phi\colon\widetilde M\rightarrow\widetilde M$ as
$\phi=R_{\gamma^{-1}}$. From \Xtrans,
     $$
     (\phi)_* V=X_J T^J{}_I C^I.
     \eqn\tmp
     $$
In general, a diffeomorphism between two manifolds induces an
isomorphism between the tensor fields (of the same rank) on each
manifolds. By such an isomorphism $\phi^\#$ induced from $\phi$,
\LiedOmega\ is mapped into
     $$
     {\cal L}_{\phi^\# X_I}(\phi^\# \Omega)=C_I\,(\phi^\# \Omega).
     $$
By noting that $\phi^\# \Omega=(\phi^{-1})^* \Omega=\Omega$ since
$\gamma\in\Gamma\subset\hbox{Isom}(\widetilde M,\widetilde g)$ and
that $\phi^\# X_I=\phi_* X_I=X_J T^J{}_I$ from \Xtrans, we obtain
     $$
     C_I=C_J T^J{}_I,
     \eqno\eq
     $$
from which one can easily show that \tmp\ reduces to
     $$
     (\phi)_* V=X_I C^I=V.
     \eqno\eq
     $$
This means that $V$ is invariant under the covering transformation,
so it is globally defined on $M$. Thus, for a globally  defined
vector field $V$ on $M$, the equation \VLiedOmega\ holds. It then
follows that there exists a 1-parameter family of transformations
$f_t:(M,g)\rightarrow(M,g)$, generated by $V$, such that
$(f_t)^*\Omega=\exp\{t\,(C_I C^I)\}\,\Omega$. This implies that
     $$
     \int_M\,\Omega=\int_M\,(f_t)^*\Omega
     =\exp\{t\,(C_I C^I)\}\,\int_M\,\Omega,
     \eqno\eq
     $$
which contradicts with the finiteness of the volume of closed $M$.
This completes the proof of the proposition.
The admissible class~B type \B 3, \B 5 and \BA 7{A\not=0} can be
compactified in such a way that \condition\ does not hold.

In particular, there is no class~B model whose spatially homogeneous
sections are closed and simply homogeneous. For there globally exists
an invariant basis on a simply homogeneous manifold, but it implies
that $\Gamma\subset G_3$. This particular case can be shown more
easily. A simple calculation shows that the following identity holds
for an integral of a spatial divergence, %
     $$
     \int_M d^3x\ \partial_i(\vert\chi\vert\,X^i_I)=
     C_I\int_M\,\Omega,
     $$
where $\vert\chi\vert$ is a determinant for the matrix $\chi_i^I$.
Since the invariant basis is globally defined in this case, it gives a
contradiction for a closed class~B model. In fact, such a
spatial divergence arises as typical boundary terms in the variation
of an action for general Bianchi models. For class~B, open or closed,
it implies the absence of a general scheme of action principle
where one imposes the spatial homogeneity before taking variation of
the action, as is well known \refmark\MT.

Finally, we would like to comment on how such study of closed
Bianchi models can be used to investigate global dynamics in
(3+1)-dimensional gravity.

The canonical formalism of general relativity treats a space-time as
a dynamical deformation of a spacelike hypersurface and a time
evolution of its three-dimensional geometry. It is important to know
what are the dynamical degrees of freedom of three-dimensional
geometry and how they evolve in time according to the classical
equations of motion, in order to understand both classical and quantum
aspects of gravity. (2+1)-dimensional gravity serves as a toy model
for studying these things \refmark\M \refmark\HN \refmark\IO. In
(2+1)-dimension, there are no local gravitational-wave modes but one
has global modes, which are related to the so-called \lq\lq
moduli\rq\rq\ of a closed 2-manifold. By global dynamics in gravity,
we mean such dynamical degrees of freedom that is closely related to
the topology of a spatial manifold.

When one proceeds to consider the real life of (3+1)-dimensional
gravity, it should be first made clear what is global dynamics in
(3+1)-dimension. It would not be easy to define global deformation of
three-dimensional spatial manifold in the full dynamics of
(3+1)-dimensional gravity. As a first step to understand it, one can
restrict one's attention to spatially homogeneous space-times. As we
have observed, each spatially homogeneous space is a locally
homogeneous manifold which can be recovered from its universal
covering space as a quotient space. This is in the same situation as
one could define \lq\lq moduli\rq\rq\ of a closed 2-manifold. But
since the universal covering space is not always a maximally
symmetric space as we saw in Thurston's theorem, it is necessary to
distinguish anisotropy from moduli degrees now. Then, one can define
moduli of a locally homogeneous 3-manifold by examining the possible
quotient spaces of each type of universal covering spaces (this is in
progress by another group \refmark\HKT). It is noted that this
viewpoint concentrates only on the geometry of spatial sections so
that it may need a certain modification in the space-time
construction.

As the next step, one can take into account a deviation from the
locally homogeneity presented here. One thing is to include a small
fluctuation of local gravitational-wave modes in an appropriate
perturbation scheme, or to examine the coupling of the global modes
with some matter fields. Another is to consider \lq\lq topological
inhomogeneity\rq\rq. It is conjectured by Thurston \refmark\TP\ that
any compact 3-manifold has a kind of prime decomposition, each element
of which admits one geometry of the eight classes described. Then it
is an interesting problem what is a possible geometry on a manifold
that is a connected-sum of such prime manifolds and how one can
glue two locally homogeneous geometries of different types along a
junction surface in between.

\vfill\eject
\par \penalty-400 \vskip\chapterskip
   \spacecheck\referenceminspace \immediate\closeout\referencewrite
   \referenceopenfalse
   \line{\fourteenrm\hfil REFERENCES\hfil}\vskip\headskip
   \input reference.aux

\end


\documentstyle[12pt]{article}

\advance\voffset by -2cm
\addtolength{\textheight}{4cm}
\addtolength{\oddsidemargin}{-1cm}
\addtolength{\textwidth}{3cm}

\def\bm#1{\mbox{\boldmath $#1$}}
\def\RF{\bm{R}}

\pagestyle{empty}

\begin{document}

\begin{table}[t]
\begin{tabular}[t]{lll}
\hline\hline
Bianchi Type & Geometry & Isotropy \\
\hline\hline\\
{\bf Class A }&&\\
&&\\
I    & $R^3$ 		& $e$   		\\
     & $E^2\times\RF$&  $SO(2)$ 		\\
     & $E^3$ 		& $SO(3)$ 	        \\
II   & Heisenberg G	& $e$ 			\\
     & Nil 		& $SO(2)$ 		\\
VI(0)& Sol=E(1,1)	& $e$ 			\\
VII(0) & $E(2)$ 	& $e$ 			\\
     & $E^3$ 	    	& $SO(3)$ 		\\
VIII& $\widetilde{SL_2\RF}$& $e$	        \\
     & $\widetilde{SL_2\RF}$&$SO(2)$            \\
IX   & $SU(2)\approx S^3$ & $e$ 		\\
     & $SU(2)\approx S^3$ & $SO(2)$ 	        \\
     & $S^3$         & $SO(3)$		        \\
&&\\
\hline
\end{tabular}
\begin{tabular}[t]{lll}
\hline\hline
Bianchi Type & Geometry & Isotropy \\
\hline\hline\\
{\bf Class B}&&\\
&&\\
III=VI(1)  & $H^2 \times E^1$ & $SO(2)$ 	\\
     & $\widetilde{SL_2\RF}$  & $SO(2)$         \\
IV   & --- &                                    \\
V    & $H^3$ 		&  $SO(3)$ 	        \\
VI(A$\not=$0,1)  & --- &                        \\
VII(A$\not=$0) & $H^3$ 	& $SO(3)$ 	        \\
&&\\
&&\\
&&\\
&&\\
&&\\
&&\\
&&\\
&&\\
\hline
\end{tabular}
\caption{Geometry of the compactified Bianchi type}
\end{table}

\bigskip

\bigskip

\begin{table}[b]
\begin{center}
\begin{tabular}{lll}\hline\hline
Thurston Type & \multicolumn{2}{c}{Bianchi Type}\\
& Class A & Class B\\ \hline\hline
     	&	    	        &		\\
$E^3$ 	& 	I, VII(0) 	& 		\\
$H^3$ 	& 			& V, VII(A$\not=$0) \\
$S^3$ 	& 	IX 		& 		\\
$S^2\times E^1$ & --- 		& --- 		\\
$H^2\times E^1$ & 		& III=VI(1) 	\\
$\widetilde{SL_2\RF}$& VIII	& III=VI(1)	\\
Nil 	& 	II		& 		\\
Sol 	& 	VI(0)		& 		\\
	&			&		\\
\hline
\end{tabular}
\end{center}
\caption{The correspondence between the Thurston type
and the Bianchi type}
\end{table}

\end{document}